\documentstyle[12pt]{article}
\normalsize

\newcounter{sxn}

\newcounter{axn}

\def\br{\begin{thebibliography}{abc}}
\def\er{\end{thebibliography}}

\date{}

\begin{document}
\bibliographystyle{unsrt}
\footskip 1.0cm
\thispagestyle{empty}
\begin{center} {\Large \bf Theoretical Physics Institute\\
University of Minnesota}\end{center}
\vspace*{3mm}
\begin{flushright}
TPI-MINN-92/23-T\\
NSF-ITP-92-139i\\
September 1992\\
\end{flushright}
\vspace*{6mm}
\centerline {\Large NUMERICAL STUDY OF THE LOWEST ENERGY CONFIGURATIONS}
\vspace*{3mm}
\centerline {\Large FOR GLOBAL STRING-ANTISTRING PAIRS}
\vspace*{10mm}
\centerline {\large Ajit Mohan Srivastava \footnote{Present address:
Institute for Theoretical Physics, University of California,
Santa Barbara, CA 93106, USA} }
\vspace*{5mm}
\centerline {\it Theoretical Physics Institute, University of Minnesota}
\centerline {\it Minneapolis, Minnesota 55455, USA}

\vspace*{10mm}

\baselineskip=18pt

\centerline {\bf ABSTRACT}
\vspace*{5mm}

 We investigate the lowest energy configurations for  string - antistring
pairs at fixed separations by numerically minimizing the energy.
We show that for separations
smaller than a critical value, a region of false vacuum develops in
the middle due to large gradient energy density.
Consequently, well defined string - antistring pairs do not exist for
such separations. We present an example of vortex - antivortex
production by vacuum bubbles where this effect seems to play a
dynamical role in the annihilation of the pair. We also study the
dependence of the energy of an string-antistring pair on their separation
and find deviations from a simple logarithmic dependence for small
separations.

\newpage

\newcommand{\be}{\begin{equation}}
\newcommand{\ee}{\end{equation}}

\baselineskip=18pt
\setcounter{page}{2}

\vskip .1in

\centerline {\bf 1. INTRODUCTION}
\vskip .1in

  Production and interaction of topological defects has been the subject of a
large number of investigations. Existence of such defects in various
condensed matter systems has been known for long time. Topological
defects can also arise due to the spontaneous symmetry breaking in the
early Universe. There are essentially two different sorts of
mechanisms for the production of topological defects. They could be
produced due to spatial fluctuations in the vacuum degrees of freedom
of the Higgs field [1], or they could
be produced directly due to energy fluctuations [2]. These two processes
are unrelated and depending upon the situation, one may dominate over
the other. (This is especially true for the global case. For the gauged
case, the basic picture of [1] has been recently reanalyzed, see [3].)
For example in the context of the early Universe, the number
of global defects which exist in the broken phase as the universe cools
through a phase transition will generally be dominated by the first
kind of process as the number of thermally produced defects will be
exponentially suppressed at low temperatures. On the other hand,
if we were to consider say the heating of a sample of liquid crystal
towards the transition temperature, then the production of
defect-antidefect pairs may be dominated by the energy fluctuations.

 When global defects are produced due to energy fluctuations then
due to the fact that the topological
quantum numbers are conserved in local fluctuations one needs
to consider the production of  defect-antidefect
pairs due to local fluctuations in the energy. For 3+1 dimensions,
one may consider monopole-antimonopole pairs, or small loops of strings
whereas for 2+1 dimensions one may consider the production of vortex-
antivortex pairs. Typically one will expect that a pair where defect and
antidefect are far separated
will be suppressed, first due to the local nature of the energy
fluctuation and secondly, for global defects, the
energy of a far separated pair is larger leading to additional suppression.
One needs to, therefore, consider the production of a defect-antidefect pair
with a given separation between the defect and the antidefect.
Further, for a pair with a given defect-antidefect separation, various kinds
of field configurations will exist and the one with lowest energy will
dominate. As a defect-antidefect configuration is not a solution
of static equations of motion, the above discussion amounts to finding
the lowest energy configuration with the constraint that the distance
between the defect and the antidefect be held fixed. We would like
to mention here that similar kind of
constrained minimization has been recently used for instanton-anti-instanton
pairs in the context of determining multiparticle production
cross-sections at high energies, see [4].
It may be interesting to see if the approach we develop here can be used for
those cases as well.

 In this paper, we have carried out the numerical minimization of energy
for  global U(1) strings. We study a pair of parallel string
and antistring, and find the lowest energy configuration
for the pair by keeping the separation fixed. We find that such
a configuration always does not exist. This is because when the separation
between the string and the antistring is less than a certain critical value
(which turns out to be about 8 - 9 times the inverse Higgs mass) then the
gradient energy density in the intermediate region becomes very large
such that it becomes possible to produce a pair of string-antistring
in the intermediate region which annihilates, cancelling the initial
winding numbers and thereby lowering the gradient energy. This suggests
that the production of vortex-antivortex pairs separated by  distances less
than a certain critical value (or string loops with small diameters) will
be suppressed in a thermal production
process from what one might have expected. This is
because the only way in which such a pair can have a distinct
identification of a vortex and an antivortex is by having highly excited
configuration. We also study the dependence of the energy of the
string-antistring pair on their separation and find that, for small
separations, the energy varies as proportional to $(ln(R))^\alpha$, R being
the separation between the string and the antistring, and $\alpha$
numerically found to be $\simeq$ 0.32.

  An interesting implication of the existence of a critical separation
between the string-antistring pair is for the case of
string-antistring annihilation process. Generally in such studies
the string and the antistring move with large velocities. Even if
they start at rest, due to long range forces between global strings,
they collide with large velocities. In most of such cases then,
the above mentioned effect, namely, the intermediate region getting
large energy density and leading to the annihilation, does not play
any role. However, it may play the dominant role in the annihilation
process in some special cases where the string and antistring are very close
and the relative velocity is not too large
as we will show in an example of the
production of a vortex-antivortex pair by vacuum bubbles. The pair
annihilation in that case proceeds by the evolution of the intermediate
region into false vacuum and winding number disappears much faster
than it could even if the vortices collided with speed of light.

 We first discuss the constrained minimization of the energy of a
string-antistring configuration in Sec. 2. The case of
vortex-antivortex production by bubbles and subsequent
annihilation is discussed in Sec. 3 and conclusions are presented in Sec. 4.

\newpage

\vskip .3in

\centerline {\bf 2. LOWEST ENERGY STRING-ANTISTRING CONFIGURATIONS}
\vskip .1in

We consider a model with a single complex scalar field
where strings arise due to the spontaneous breaking of a global U(1)
symmetry

$$L = {1 \over 2}(\partial_\mu \Phi^*)(\partial^\mu \Phi) -
{\lambda^2 \over 4}(\Phi^* \Phi - \eta^2)^2 \eqno(1) $$

\noindent where $\Phi = \Phi_1 + i\Phi_2 = \phi e^{i\theta}$.
We will use the natural system of units with $\hbar$ = c = 1.
All distances will be measured in the units of $(\eta \lambda)^{-1}$,
energy density in the units of $\lambda^2 \eta^4$ and $\phi$
in units of $\eta$. We will use  $\lambda = \eta = 1$ and therefore
all these units are equal to 1.

  The above model admits string solutions in the broken symmetry phase
which, for a winding number one
string parallel to the z axis, can be written as

 $$ \Phi(r,\theta) = \phi(r) e^{i\theta} \eqno(2) $$

\noindent where $r$ and $\theta$ are respectively radial and
azimuthal coordinates in the x-y plane. For the static case,
$\phi$ satisfies the following  equation

 $$\phi^{\prime \prime} + {\phi^\prime \over r} - {\phi \over r^2}
 - \lambda^2 \phi (\phi^2 - \eta^2) = 0 \eqno(3) $$

  The energy (per unit length, assuming z symmetry) associated with
a given configuration is

 $$ E = \int d^2x [ {1 \over 2} |\bigtriangledown \Phi |^2
    + {\lambda^2 \over 4} (\Phi^* \Phi - \eta^2)^2 ] \eqno(4) $$

  Our minimization procedure consists in starting with a given
field configuration  on a two dimensional lattice and then
varying the field configuration at each lattice site. Since we are
attempting to find the lowest energy configuration, we allow fields
at the boundary to vary as well. We have tried
out various minimization techniques and have found that over relaxation
is very efficient for our case. This consists in first determining
the most favorable fluctuation in $\Phi$ at a given site by fluctuating
$\Phi$ there and considering the change in the energy density. The most
suitable fluctuation corresponds to the minimum of the parabola which
passes through these values of energy densities (corresponding to
fluctuated values of $\Phi$). Then the
actual change in $\Phi$ is taken to be larger (by a certain factor)
than this most suitable fluctuation. We have found that changing this
factor in the range of 0.2 - 0.9 worked best for our case. Computations
were carried out on Cray-2 and Cray X-MP computers at the Minnesota
Supercomputer Institute.

  We have tested our minimization code by finding the configuration of
a single string. We take the form of $\Phi$ as given in Eq.(2) and
prescribe some initial function for $\phi(r)$. We then minimize the
energy (Eq.(4)) and determine $\phi(r)$ which gives the lowest
energy configuration. We have found that even if the initial profiles
for $\phi(r)$ prescribed are very different (for example we have
tried out triangular form for $\phi(r)$) , after about 200
iterations, $\phi(r)$ converges to the exact solution as obtained
from Eq.(3). In Fig. 1 we have given the resulting $\phi(r)$. For this case
we start with $\phi(r)$ given by $\phi(r) = \eta (1 - e^{-r/\delta})$
with $\delta$ chosen to be 0.4. We choose such a $\delta$ so that the
initial profile is very different from the correct one. This initial
profile is shown in Fig.1 by the dotted curve. The correct
solution for $\phi(r)$ is obtained by numerically solving Eq.(3)
using a Runge- Kutta algorithm of fourth order accuracy. The solution is
shown in Fig. 1 by the solid curve. It is
known [5] that for large $r$ the leading terms in a power series
expansion of $\phi(r)$ are given by

 $$ \phi(r) = \eta ~(1 - {1 \over 2r^2}) \eqno(5) $$

 By fitting the large $r$ region of the solution (solid curve in
Fig. 1)  we find
the exponent of $r$ in Eq.(5) to be 1.996 and the coefficient
of 1/$r^2$ to be 0.503.
Starting with $\phi(r)$ as given by the dotted curve in Fig.1,
after about 200 iterations we obtained the dashed curve in Fig.1
which is extremely close to the correct solution showing the efficiency
of our energy minimization code.

  We now continue to determine the lowest energy string-antistring
configuration at a fixed separation. First, it is helpful to make a
rough estimate of the gradient energy density contained in the
region between the string and antistring.  For this purpose we take the
ansatz which we used in an earlier work for the case of global string loops
[6].  (In [6], the core energy was neglected and only the gradient energy
outside the core was considered. It does not matter, however, as the only
thing we need from [6] is the extent of
the region in which the gradient energy is concentrated at the midpoint
of the loop.) It was assumed in [6]
that all (or most) of the gradient energy is contained within the region
bounded by two paraboloids. The gradient energy density is smallest
near the center of the loop where the distance between either of the
paraboloids from the center of the loop (called Z$_0$ in [6]) is largest.
Taking Z$_0$ as a variational
parameter, it was found in [6] that Z$_0 = {\sqrt{3} \over 4} $ D where
D is the diameter of the loop. As the loop was assumed to be azimuthally
symmetric in [6], we can take its intersection by a plane normal to the loop
and passing through it's center. This gives us the Higgs phase distribution
for a parallel string-antistring
pair separated by a distance D such that all of
the gradient energy is concentrated within two outermost parabolas,
see Fig. 2. The distance between the two parabolas at the midpoint is
2Z$_0$.  The gradient energy density is smallest
near the midpoint and if in that region it becomes larger than the
false vacuum energy density, we will expect that well defined string-
antistring configuration does not exist anymore. This will happen when

 $$ {1 \over 2} \eta^2 ({2\pi \over 2Z_0})^2 >
  {\lambda^2 \over 4} \eta^4 \eqno(6). $$

 Since Z$_0 = {\sqrt{3} \over 4} D$, this implies

 $$ D < \sqrt{2 \over 3} ~~ {4\pi \over \lambda \eta} \eqno(7) $$

 With our choice of parameters ($\eta = \lambda = 1$) this implies
that when D $<$ 10.3, a well defined string - antistring pair will not exist.
We will see later that our numerical results confirm this estimate to
a reasonable accuracy where we find the critical separation to be
$\simeq$ 12.0. [Actually the gradient energy density is higher near the
strings which is where the false vacuum first develops, as we will see later.
This may account for somewhat larger value of the critical separation
we find.] This suggests that the ansatz used in [6]
correctly represents the concentration of the gradient energy between
the string-antistring pair (even though it does not correctly describe
the field configuration near the cores of the individual strings).

  We now consider the energy minimization for a string-antistring
pair with their separation held fixed. All along we will take the
string and antistring to be along z axis and only present
their profiles in the x-y plane. Our results are thus also valid for
vortex-antivortex pairs in 2+1 dimensions. We take the initial profiles of
the string and the antistring to be the ones obtained by numerically
solving the equations of motion (the solid curve in Fig.1). To give
configuration of the pair we take the product ansatz [5]. If the string
center is located at $\vec r_1$ and the antistring center at $\vec r_2$ then
$\Phi$ for the pair is given by

 $$\Phi_{pair}(\vec r) = {1 \over \eta} ~\Phi_{string}(\vec r - \vec r_1)
\Phi_{antistring}(\vec r - \vec r_2)\eqno(8)$$

We hold a string (or antistring) fixed by fixing the field configuration
at few nearby points on the lattice.
Only fixing $\Phi$ at the center of the string  does not work since
it only fixes the magnitude of $\Phi$ without fixing it's winding number
and it becomes energetically favorable for the pair
to let the winding slip out of the fixed centers and
annihilate it in the middle. We consider the center of the string
(antistring) to lie at the midpoint of an elementary plaquette and then
hold $\Phi$ fixed at the four corners of this plaquette thereby fixing
the winding number. [We have also
tried fixing $\Phi$ in somewhat larger region by
considering the center of the string at a lattice site and holding
$\Phi$ fixed at the corners of all four plaquettes which have one vertex
common with the center of the string. The results are essentially the same.]
We use 400 $\times$ 400 lattice with the physical size of 80 $\times$ 80.
The choice of the lattice size was governed by the fact that if a string
is too close to the boundary then it becomes energetically favorable for it
to have its gradient energy concentrated towards the boundary. To avoid this
``boundary" effect we considered lattice size such that the separation between
the string-antistring is smaller than the distance of either of them from
the lattice boundary.

 Fig. 3a shows the energy density plot for a string-antistring pair
with separation equal to 14 units. In order to show string-antistring
configuration clearly we will always plot only the central portion of the
lattice with physical size of 20 $\times$ 20. Minimization is carried out
until the energy is almost stationary. The energy density plot for the
final configuration is shown in Fig. 3b. The strings get little squeezed
towards the middle region and energy density gets little peaked
near the centers of the strings (due to holding the strings fixed).
The profile of $\phi$ does not change much (except
little squeeze towards middle) and we do not show it here. We therefore see
that for this separation (as well as for larger separations)
it is possible to find the lowest energy configuration for a well defined
string-antistring pair. We then continue
the minimization but now by letting the strings move. String-antistring
successively approach each other and annihilate as shown by Fig.3c
and Fig.3d.

 We now consider a pair with smaller separation. Fig. 4a shows the
energy density plot for a pair with separation equal to 12 units. Fig. 4b
is the plot of $\eta - \phi$ for this pair. We carry out the minimization
but in this case we do not achieve any stationary value of the energy
and the energy keeps decreasing. We show the plots at various intermediate
stages. As the minimization proceeds, the energy density becomes highly
peaked near the centers of the strings as shown in Fig. 4c.
(If the peak becomes higher than 1.2 then we truncate it for
plotting convenience.) Again this happens because we have held
$\Phi$ near the string center fixed while $\Phi$ in the neighboring region
is distorting to keep energy lowest. Fig. 4d shows the plot of
$\eta - \phi$ at the same stage (as in Fig. 4c)  showing clearly that
the whole profile of string and the antistring has squeezed towards
the intermediate region and two peaks have developed in the
intermediate region. These peaks are actually a pair of string-antistring
which get created in the middle region.
This happens as the Higgs phase gradient
energy concentrates more and more in the middle region and given that the
gradient energy is highest near the strings, it becomes favorable for
$\Phi$ to develop zeros in those regions. The distortion near the string
centers is so large that the winding number near the original string centers
disappears (due to finite lattice spacing) and is carried by the new pair.
Figs.4e-4g show the successive stages where the intermediate
peaks (the new string-antistring pair) come towards each other
and annihilate. The two remaining peaks in Fig. 4g are due to holding
the string centers fixed. When we let $\Phi$ vary everywhere then these
peaks quickly decay away as shown in Fig. 4h.
Fig. 4i is the  Higgs phase plot at the initial stage
corresponding to Fig. 4a and shows the windings of the string and antistring.
Fig. 4j corresponds to the stage as in Fig. 4g showing clearly that all
the windings have disappeared. Exactly near the initial locations of the
centers of string-antistring the winding numbers are still held fixed. However,
due to finite lattice spacing, the continuity can be broken if it is
energetically favorable which is what happens here. The peaks
in the energy density (e.g. Figs. 4d and 4g) are precisely due to this
rapid change in the Higgs phase near the string centers (these peaks
disappear in Fig. 4h when $\Phi$ is not held fixed anymore).

 Exactly the same behavior, as we observed above for separation equal to 12,
was observed for separations smaller than 12 as well. For separation equal to
13 the situation was not very clear as the energy seems to become
stationary after a large number of variational steps. However, the
profile of $\phi$ develops edges near the centers of the strings
(somewhat similar to the one in Fig. 4d, but the second peak being
extremely close to the original string). Separation equal to 13 thus
seems like the border line case. Clear distinct behavior is observed
between separation $\ge$ 14
case (where the string and antistring remain almost
unchanged, except little squeeze towards the middle region) and
separation $\le$ 12
case where it is not possible to hold string and antistring
separately. Note that this is in good agreement with the estimate
of the cutoff separation in Eq.(7).
What this means is that the lowest energy configurations
for separations less than 12 do not resemble in any way to string-antistring
configurations. In a thermal production one will generally expect that pairs
with smaller and smaller separations will be more and more abundant
(especially for vortex-antivortex pairs).
This does not seem to be the case though as pairs with distances less than 12
(for general parameters this cutoff distance will be of order of
what is given in Eq.(7)) will exist only if they are highly excited.
One may thus expect the number density of string-antistring pairs
to not keep increasing as a function of separation and
(at best) level off at a cutoff separation. For string loops these results
suggest that, again due to the concentration of the gradient energy in the
inside of the loop, there will be a critical diameter such that well defined,
lowest energy, configurations for loops with smaller (fixed) diameters will
not exist.

  We now study the dependence of the energy of the string-antistring pair
on their separation. We found the lowest energies for separations $R$
equal to 14, 16, 20, 24, 28 and 32. For larger values of $R$ (actually
even for $R$ = 32), the strings are closer to the boundary and the values
of energy are affected by the boundary cutoff (as we mentioned earlier).
For large $R$, $E$ is supposed to vary in the following manner [5],

 $$E = A ~ ln(R/{\lambda})   \eqno(9) $$

\noindent  with $A$ of the order of $\pi$ and $\lambda$ of the order of
inverse Higgs mass. (Energies of the cores can be absorbed by redefining
$\lambda$).

Fig. 5 shows the variation of the energy $E$ obtained by our minimization
code, as a function of $ln(R)$. Small squares show the values of $E$ at
the above mentioned values of $R$. There is a clear deviation from a
straight line. We have fitted the two different segments of straight lines
(shown by the two dashed line segments) to these points. For small $R$,
best fit to first three points gives $A \simeq$ 6.20 and $\lambda \simeq$ 0.73,
while for large $R$, best fit to last three points gives
$A \simeq$ 4.35 and $\lambda \simeq$ 0.17. Theoretically expected value of
$\lambda$ is $\simeq$ 0.71 (corrections in the definition of $\lambda$ due
to core energy are small). Although the value of $A$ obtained by fitting
points at large R is closer to theoretically expected value, the value of
$\lambda$ is not. The increase in the slope of the energy curve
for smaller values of $R$ may be related to the existence of the critical
separation. However, we again emphasize that values of $E$ at large
$R$ are affected by boundary cutoffs which may affect the slope.
We also attempted to fit a curve
where $A$ in Eq. (9) was replaced by $A R^\beta$. However, we could not
find a good fit  for any choice of (positive or negative)  $\beta$,
$A$ and $\lambda$.
The curve which describes a good fit to all six points is given by

 $$ E = A (ln(R/{\lambda}))^{\alpha}  \eqno(10)$$

\noindent with $A \simeq 19.3, \lambda \simeq 6.0$ and $\alpha \simeq 0.32$.
This is shown by the solid curve in Fig. 5. Larger values of $\alpha$ do
not fit all the points so well.
However, we would like to mention that the value of $\alpha$
here may be affected by boundary effects as we mentioned above.


\vskip .3in

\centerline {\bf 3. VORTEX-ANTIVORTEX PRODUCTION BY BUBBLES}
\vskip .1in

 As we had mentioned earlier, the above results also have implications for
string-antistring annihilation processes. However in most such cases the
intermediate region does not have time to evolve due to large velocities
of the strings. One has to then find special cases where the string-
antistring get created very close to each other and
with  small relative velocities.
We now present an example where this is what seems to happen and
the annihilation of string-antistring pair appears to be
completely dominated by the sort of behavior
we discussed above. We study a first order phase transition case
in 2+1 dimensions where the vortices are produced by the collision
of vacuum bubbles (see [7], for details). The Lagrangian density is given by

  $$ L = {1 \over 2}\partial_\mu \Phi^*\partial^\mu \Phi
  - {\lambda \over 4} \phi^2
(\phi - \phi_0)^2 + {\lambda \over 2} \epsilon \phi_0 \phi^3 \eqno(11)$$

\noindent where $\phi$ is the magnitude of $\Phi$ ($\Phi = \phi e^{i\theta}$).
There is a metastable vacuum at $\phi$ = 0 and the true vacuum is
at $\phi = \eta^\prime$ where $\eta^\prime$
is the vacuum expectation value of $\Phi$
which spontaneously breaks the U(1) global symmetry leading to the existence
of global strings. Following results are for parameter choices
$\epsilon = 0.1, \phi_0 = 4.0$
and $\lambda = 4.0$. We will measure spatial and temporal coordinates
in terms of the inverse Higgs mass which for these values of parameters
is equal to 0.12 (we continue to use the natural system of units).

 The phase transition in this case proceeds by the nucleation of critical
vacuum
bubbles whose profile is given by the solutions of the Euclidean equations
of motion. The actual details of this are not relevant here and
we refer the reader to Ref. [7]. It was shown in [7]
that when three critical bubbles collide then depending on the values
of Higgs phases inside the bubbles,  a vortex may form at the collision
point. We had also found in [7] that vortex can form even in the
collision of two critical bubbles and a subcritical bubble; a subcritical
bubble being a small bubble which collapses (and then bounces back
before collapsing again). We had found in [7] that the vortex found
in this manner invariably escapes out of the bubbles because of the
large momentum of the walls of critical bubbles compared to the momentum
of the wall of the subcritical bubble. [Here we may mention again that
use of classical equations of motion for subcritical bubbles is
really justified only for the case of thermal production. One may thus
consider the case of thermal production and
take our critical and subcritical bubbles as just representing a class
of expanding and collapsing bubbles respectively.]
 We had also found in [7] that if the Higgs phase distribution
is asymmetric in the three colliding critical bubbles then in order to
minimize the gradient energy, the
vortex develops large velocity towards the direction of larger
phase gradient.

 We use these results now and consider the collision
of two critical bubbles and a subcritical bubble such that most
of the phase gradient energy is concentrated towards the critical bubbles
(from the collision point) which should then counter the effect
of large momentum of critical bubble walls. The idea being that
this way one may be able to stop the vortex from escaping out
of the bubble. What we find however is that although the vortex
itself does not escape out of the bubble, an antivortex gets
created at the bubble wall which moves in and annihilates this
vortex. The whole thing being consistent with the fact that
the final field configuration in space is the one given only by the
two critical bubbles.

 Fig. 6a is the plot of $\eta^\prime - \phi$ and
shows the initial profiles of bubbles. Left one is the
subcritical bubble while the two on the right are the critical
bubbles. In the plots of $\eta^\prime - \phi$, the x axis will be  from top
to bottom and the y axis from left to right.  Fig. 6b shows the initial
distribution of the Higgs phase $\theta$ for these bubbles. Starting
with the bubble with smallest value of y (which is the subcritical bubble)
and going counter clockwise, the values of $\theta$ are respectively
180$^0$, 100$^0$, and 260$^0$. $\theta$ for the subcritical bubble
will flip to $\theta$ = 0 after the bubble collapses and bounces back,
see [7]. Thus the variation of $\theta$ is maximum between the top two
(critical) bubbles. Fig. 6c shows the situation at t = 20.2
when all three bubbles have coalesced and a vortex
is formed in the collision region.
Figs. 6d - 6g show closeup of the region of the vortex.
Fig. 6e shows an antivortex separating from the bubble wall at t = 24.37
(which will become clear when we show plots of Higgs phase). We see
clearly that the region between the vortex and the bubble wall
evolves to the false vacuum. Vortex-antivortex  annihilate each other
by t = 24.76 as shown in Fig. 6f and finally decay away, see Fig. 6g.

 Let us now follow the annihilation process by following the plots of the
Higgs phase. Fig. 7a-7c are the plots of the region containing the vortex.
Fig. 7a shows only one vortex whereas Fig. 7b shows the presence of
a well defined pair of vortex-antivortex near X $\simeq$ 56.0
(corresponding to the plot in Fig. 6e). Vortex being at
y $\simeq$ 49.0 and the antivortex at  y $\simeq$ 43.0.
[The lengths of vectors
in these figures are large for large $\phi$ and vectors are not plotted
where $\phi$ is extremely small. Fig. 7b therefore shows that vortex
and the antivortex are clearly separated by region where $\phi$ is
different from zero.] Fig. 7c shows the case when the vortex
and antivortex have disappeared. If this annihilation had proceeded
by the vortex and antivortex moving towards each other, it will imply
a relative velocity of about 14 times the speed of light,
clearly an absurd number. What instead
happens here is that the region in between the vortex and antivortex
evolves to $\phi$ = 0 and essentially dissolves the vortex and antivortex.
This is the same sort of behavior we had observed in the variational study
discussed in Sec.2.

\vskip .3in
\newpage

\centerline {\bf 4. CONCLUSIONS}
\vskip .1in

  We emphasize again that the phenomena we discuss here, namely
the region in between the string-antistring playing a crucial role
in the annihilation process, will generally be obscured in the
studies where the string and antistring move with large velocities.
In those cases, the annihilation will proceed by the string and antistring
approaching each other. Under very special situations, such as
the one discussed above for the case of bubbles, it may happen that the
antistring just gets created close to the string (with little
relative velocity) and one may be able to observe such effects.
Our results for the lowest energy configurations of
string-antistring pairs (vortex-antivortex pairs in 2+1 dimensions)
show the existence of a cutoff separation and suggest that if
defect-antidefect pairs were thermally produced then
their number density should not keep increasing
and at best may  flatten out for separations smaller than a critical value.
For string loops our results imply that loops with diameters less than
a critical diameter may be suppressed compared to naive expectations.
We have also studied the variation of the energy of a string-antistring
pair on the separation $R$ and find that, for small $R$, it deviates from
a simple logarithmic  dependence. These results should have consequences
for defect production in phase transitions (such as the one studied
in [8]) and may be testable in condensed matter experiments.
All these qualitative features should clearly
exist for other global defects as well, such as global monopoles etc.

\vskip .3in

\centerline {\bf 4. ACKNOWLEDGEMENTS}
\vskip .1in

I am very grateful to M.B. Voloshin for his suggestions on the over
relaxation method.
This work was supported by the Theoretical Physics Institute at the
University of Minnesota, by the U.S. Department of Energy under
contract number DE-AC02-83ER40105, by a grant from the Minnesota
Supercomputer Institute and by the National Science Foundation
under Grant No. PHY89-04035.

\vskip .3in
\newpage

\centerline{$\underline{\it REFERENCES}$}

\begin{enumerate}

\item T.W.B. Kibble, J. Phys. A9, 1387 (1976).

\item F. A. Bais and S. Rudaz, Nucl. Phys. B170, 507 (1980);
M. B. Einhorn, D. L. Stein, and D. Toussaint,
Phys. Rev. D21, 3295 (1980).

\item S. Rudaz and A. M. Srivastava, University of Minnesota preprint,
TPI-MINN-92/20-T, UMN-TH-1028/92, (August 1992).

\item V. V. Khoze, J. Kripfganz and A. Ringwald, Phys. Lett. B275, 381 (1992)
and references therein.

\item E.P.S. Shellard, Nucl. Phys. B283, 624 (1987).

\item C. Rosenzweig and A. M. Srivastava, Phys. Lett. B222, 368 (1989).

\item A.M. Srivastava, Phys. Rev. D45, R3304 (1992); Phys. Rev. D46,
1353 (1992).

\item M. Alford, H. Feldman and M. Gleiser, Phys. Rev. Lett. 68, 1645 (1992).
\end{enumerate}

\vfil
\eject
\vskip -.1in
\centerline {\bf FIGURE CAPTIONS}

 Figure 1 : Solid curve shows the string profile obtained by solving
equations of motion. Dotted curve is the initial profile used for
minimization of energy and dashed curve gives the string profile after
the minimization is completed.

 Figure 2 : Two small circles denote cross-sections of the string and
antistring. Solid curves bound the region inside which most of the gradiant
energy of the Higgs phase is concentrated in the middle.

 Figure 3 : (a) Energy density plot for initial string-antistring pair
with separation  equal to 14.0. (b) String-antistring after energy minimization
is completed with separation fixed. (c) String-antistring pair after further
minimization with separation allowed to change. (d) Annihilation of
string-antistring.

 Figure 4 : (a) and (b) give the plots of energy density and $\eta - \phi$,
respectively for initial string-antistring pair
with separation  equal to 12.0. (c) and (d) are similar plots at an
intermediate stage of the energy minimization showing the formation of a new
string-antistring pair (which carries the winding numbers of the initial
pair) in the middle region. (e) and (f) are plots
at a later stage of the minimization showing the situation when
this new pair is about
to annihilate. (g) Field configuration after the annihilation is completed.
The two remaining peaks are due to holding the original configuration fixed
in those regions. (h) Field quickly decays away
when minimization is continued while letting $\Phi$ vary everywhere.
(i) Plot of the initial distribution of Higgs phase
showing the windings of the string and the antistring.
Higgs phase is equal to  the azimuthal angle of a vector.
(j) Final plot of the Higgs phase showing
that the winding numbers have disappeared.

 Figure 5 : Plot of energy $E$ vs. $ln(R)$. Small squares show the values
of $E$ obtained by the minimization code for various values of $R$. Two
segments of dashed lines show best fit for points at large $R$, and small
$R$ respectively. Solid curve denotes the fit $E = A (ln(R/\lambda))^\alpha$.

 Figure 6 : (a) Plot of $\eta^\prime - \phi$ showing initial configuration
of bubbles. (b) Initial distribution of Higgs phase. (c) Profile of the vortex
after bubbles have coalesced at t = 20.20. (d) Closeup of the region of the
vortex at t = 23.58. (e) Vortex at t = 24.37. Small peak separating from
the bubble wall is an antivortex. (f) Annihilation of the vortex and antivortex
at t = 24.76. (g) Decayed configuration at t = 27.5.

 Figure 7 : (a) Higgs phase plot. Vortex is located near Y $\simeq$ 49.0,
X $\simeq$ 56.0. (b) Antivortex has formed near Y $\simeq$ 43.0,
X = $\simeq$ 56.0. (c) Vortex and antivortex
have disappeared and there is no winding present.

\end{document}